\documentclass[aps,prd,twocolumn,epsf,showpacs,groupedaddress]{revtex4}

\usepackage{graphicx}
\usepackage{amsmath,amssymb,latexsym}
\usepackage{bm}

\def\lbldef#1#2{\expandafter\gdef\csname #1\endcsname {#2}}

\def\href#1#2{#2}

\newcommand{\En}{E_n}
\newcommand{\Fn}{F_n}
\newcommand{\Enu}{E_{\overline{\nu}}}

\newcommand{\Enubar}{\epsilon_{\overline{\nu}}}
\newcommand{\thnu}{\overline{\theta}_{\overline{\nu}}}
\newcommand{\cth}{\cos\overline{\theta}_{\overline{\nu}}}
\newcommand{\Fnu}{F_{\overline{\nu}}}
\newcommand{\tbar}{\overline{\tau}_n}

\newcommand{\nutau}{\nu_\tau}
\newcommand{\nue}{\nu_e}
\newcommand{\numu}{\nu_\mu}

\newcommand{\Epho}{E_\gamma}

\newcommand{\Fpho}{F_\gamma}
\newcommand{\bwide}{\begin{widetext}}
\newcommand{\ewide}{\end{widetext}}
\newcommand{\beq}[1]{\begin{equation} \label{(#1)}}
\newcommand{\eeq}{\end{equation}}
\newcommand{\ba}[1]{\begin{eqnarray} \label{(#1)}}
\newcommand{\ea}{\end{eqnarray}}

\newcommand{\HI}{\mbox{H\hspace{0.2em}{\scriptsize I}}}

\begin{document}

\preprint{
\hfil
\begin{minipage}[t]{3in}
\begin{flushright}
\vspace*{.4in}
hep-ph/0506168
\end{flushright}
\end{minipage}
}

\title{Probing Planck scale physics with IceCube}

\author{Luis A. Anchordoqui}
\email[]{l.anchordoqui@neu.edu}
\affiliation{Department of Physics,
Northeastern University, Boston, MA 02115, USA
}

\author{Haim Goldberg}
\email[]{goldberg@neu.edu}
\affiliation{Department of Physics,
Northeastern University, Boston, MA 02115, USA
}

\author{M. C. Gonzalez-Garcia}
\email[]{concha@insti.physics.sunysb.edu}
\affiliation{Y.I.T.P., SUNY at Stony Brook, Stony Brook, NY 11794-3840, USA
}
\affiliation{IFIC, Universitat de Val\`encia -- C.S.I.C., Apt 22085, 46071 Val\`encia, Spain}

\author{Francis Halzen}
\email[]{halzen@pheno.physics.wisc.edu}
\affiliation{Department of Physics, University of Wisconsin, Madison WI 53706
}

\author{Dan Hooper}
\email[]{hooper@astro.ox.ac.uk}
\affiliation{Astrophysics, University of Oxford, Oxford OX1 3RH, UK
}

\author{Subir Sarkar}
\email[]{sarkar@thphys.ox.ac.uk}
\affiliation{Theoretical Physics, University of Oxford, Oxford OX1 3NP, UK
}

\author{Thomas J. Weiler}
\email[]{tom.weiler@vanderbilt.edu}
\affiliation{Department of Physics and Astronomy, Vanderbilt University, Nashville TN 37235
}

\begin{abstract}

\noindent
Neutrino oscillations can be affected by decoherence induced e.g. by
Planck scale suppressed interactions with the space-time foam
predicted in some approaches to quantum gravity. We study the
prospects for observing such effects at IceCube, using the likely flux
of TeV antineutrinos from the Cygnus spiral arm. We formulate the
statistical analysis for evaluating the sensitivity to quantum
decoherence in the presence of the background from atmospheric
neutrinos, as well as from plausible cosmic neutrino sources. We
demonstrate that IceCube will improve the sensitivity to decoherence
effects of ${\cal O}(E^2/M_{\rm Pl})$ by 17 orders of magnitude over
present limits and, moreover, that it can probe decoherence effects of
${\cal O}(E^3/M_{\rm Pl}^2)$ which are well beyond the reach of other
experiments.

\end{abstract}


\maketitle

\section{Introduction}

Despite many decades of intense effort, a satisfactory theory of
quantum gravity is yet to see the light of day. Moreover, it is
generally thought that the quantum effects of gravity may never be
experimentally accessible because they would be manifest only at the
Planck scale, $M_{\rm Pl} \equiv \sqrt{\frac{\hbar c}{G_{\rm N}}}
\simeq 1.2 \times10^{19}$~GeV.  However, gravity, being a
non-renormalizable interaction in the language of quantum field
theory, may leave a distinctive imprint at energies much lower than
the Planck scale if it violates some fundamental symmetry of the
effective low energy theory, akin to the violation of parity in
nuclear radioactive decay at energies far below the true scale of the
responsible weak interaction~\cite{Sarkar:2002mg}. For example, if
quantum space-time has a `foamy' structure in which Planck length size
black holes form and evaporate on the Planck time scale, then there
may be a loss of quantum information across their event horizons,
providing an `environment' that can induce decoherence of apparently
isolated matter systems~\cite{Hawking:1982dj}.

The particle most sensitive to such effects would appear to be the
neutrino because oscillations between neutrino flavours are a pure
quantum phenomenon in which the density matrix, $\rho$, has the
properties of a projection operator: Tr~$\rho^2=$ Tr~$\rho=1$. A
heuristic view of decoherence induced by interactions with the virtual
black holes in the space-time foam is as follows. Since black holes
are believed not to conserve global quantum numbers, neutrino flavor
is randomized by interactions with the virtual black holes. The result
of many interactions then is to equally populate all three possible
flavors. Because black holes do conserve energy, angular momentum, and
electric and color charge (unbroken gauged quantum numbers), the
neutrino interacting with the virtual black hole does re-emerge as a
neutrino. In this connection, it has been noted
already~\cite{Lisi:2000zt} that the results from the Super-Kamiokande
atmospheric neutrino experiment~\cite{Fukuda:1998mi} and the K2K long
baseline oscillation experiment~\cite{Ahn:2002up}, interpreted in
terms of a 2-generation $\nu_\mu \leftrightarrow \nu_\tau$ flavor
transition, can probe decoherence effects with high sensitivity,
supplementing laboratory tests based on $K^0 \overline K^0$
oscillations and neutron interferometry~\cite{Ellis:1983jz}.

It has recently been suggested~\cite{Hooper:2004xr} that antineutrinos
originating in the decay of neutrons from candidate cosmic ray sources
in the Galaxy~\cite{Anchordoqui:2003vc} can provide an even more
sensitive probe. The effects of quantum decoherence would alter the
flavor mixture to the ratio, $\nu_e:\nu_\mu:\nu_\tau \simeq 1:1:1,$
regardless of the initial flavor content. Since decoherence effects
are weighted by the distance travelled by the (anti) neutrinos, this
means that if a $\overline \nu$-flux with ratio of flavors $\neq
1:1:1$ were to be observed from an astrophysical 
source, then strong constraints can be placed on the
energy scale of quantum decoherence, surpassing current bounds by over
10 orders of magnitude. However if a 1:1:1 ratio is observed, this
will not imply that quantum decoherence {\em is} responsible, since
the dominant source of the (anti) neutrinos may simply not be neutron
$\beta$-decay. In this paper, we pursue this idea further and
formulate the statistical analysis necessary for obtaining bounds on
quantum decoherence from expected future detections of cosmic
neutrinos.

In Sec.~\ref{neutrons}, we identify a candidate neutron source in the
vicinity of the Earth: Cygnus OB2. We review all existing data on the
Cygnus region and show that observed directional signals at high
energies~\cite{Cassiday:kw,Teshima:1989fc} can plausibly be ascribed
to a neutron source with an energy spectrum $\propto E_n^{-3.1}$. In
particular, because of neutron decay, the expected anisotropy is well
below limits reported by the CASA-MIA~\cite{Borione:1996jw} and
KASCADE~\cite{Maier:2003av} experiments. We summarize the estimate of
the corresponding antineutrino flux~\cite{Anchordoqui:2003vc}. In
Sec.~\ref{decoherence}, we discuss the effects of decoherence on high
energy neutrino propagation adopting the quantum dynamical semi-group
formalism, where the Lindblad operators~\cite{Lindblad:1975ef}
describe (anti) neutrino couplings to the space-time foam. In
Sec.~\ref{sec:reach}, we estimate in detail the sensitivity of the
IceCube detector~\cite{Ahrens:2003ix} to the $\overline \nu$-Cygnus
beam and its ability to constrain the effects of quantum
decoherence. We evaluate the signal-to-noise for both track and shower
events, taking into account the atmospheric neutrino background, as
well as a possible contribution to the neutrino flux from the TeV
$\gamma$-ray source TeV J2032+4130, recently discovered by the HEGRA
experiment~\cite{Aharonian:2002ij,Aharonian:2005ex} in the direction of
the Cygnus spiral arm. Armed with these event rates, we formulate the
statistical analysis required to study the sensitivity to quantum
decoherence effects. We show that IceCube can improve the sensitivity
over present probes of decoherence by 4 to 17 orders of magnitude, and
moreover, that it is sensitive to strongly energy dependent
decoherence effects suppressed by multiple powers of the Planck scale
which are beyond the reach of other experiments. Finally, in
Sec.~\ref{discussion}, we confront our results with theoretical
suggestions for quantum decoherence.

\section{
Antineutrinos from Cygnus OB2}
\label{neutrons}

\begin{figure}
\begin{center}
\includegraphics[height=9.2cm]{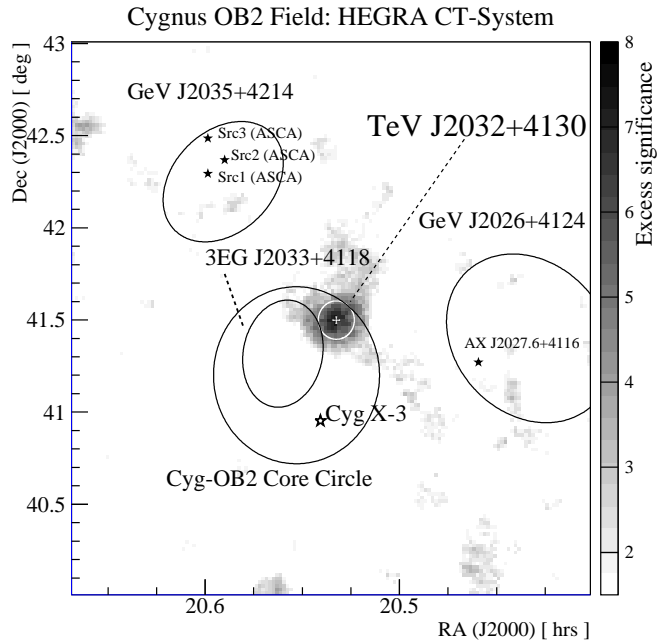}
\caption{Skymap of correlated event excess significance ($\sigma$)
from all HEGRA IACT-System data ($3^\circ \times 3^\circ$ FoV)
centered on the TeV source J2032+4130. Nearby objects are also shown:
95\% contours for 3 EGRET sources (indicated by the ovals), their
possible X-ray associated counterparts (as given in
Ref.~\cite{Roberts:2000zr}), and Cygnus X-3. The center of gravity
with statistical errors and intrinsic size (standard deviation of a
2-dim Gaussian, $\sigma_{src}$) are indicated by the white cross and
white circle, respectively. The TeV source, J2032+4130, is positioned
at the edge of the error circle of the EGRET source 3EG J2033+4118,
and within the core circle of the extremely dense OB stellar
association Cygnus OB2~\cite{Aharonian:2005ex}.}
\label{cygOB2}
\end{center}
\end{figure}

Massive star forming regions are the engines of starburst
galaxies. They generate large numbers of UV photons which ionize the
interstellar medium, as seen in microwave, radio, and H$\alpha$
recombination line emission. They are important sources of
interstellar dust heating which results in significant infrared
emission. The strong winds of their massive O stars, which should pass
through the Wolf-Rayet phase and explode as supernovae, release
considerable amounts of kinetic energy creating a rarefied hot
($\sim10^6$ K) superbubble that emits X-rays; its cavity can
eventually be discerned from observation of \HI\ and CO lines of the
interstellar gas.

Such regions are also likely sites for cosmic ray acceleration. The
massive stars synthesize considerable amounts of heavy nuclei that are
released either by stellar winds or during the subsequent supernova
explosions. Moreover, the young stellar population can create
time-correlated, clustered supernova remnants, where through
cooperative acceleration processes the energy of the accelerated
nuclei can be boosted above the ${\sim}10^{6}$~GeV cutoff of
individual remnants. Thus the usual Fermi mechanism might be able to
accelerate cosmic rays all the way up to the `ankle', where the
steeply falling ($\propto E_{\rm CR}^{-3.1}$) cosmic ray spectrum
flattens to $E_{\rm CR}^{-2.8}$~\cite{Anchordoqui:2002hs}. An
immediate consequence of this nucleus-dominance picture is the {\sl
creation of free neutrons} through the photodisintegration of the
nuclei on the intense ambient photon fields.

Independent evidence may be emerging for such a cosmic ray accelerator
in the massive star forming region Cygnus OB2, a cluster of several
thousands of hot young OB stars with a total mass of $\sim 10^4
\,M_\odot$~\cite{Knodlseder:2000vq}. At a relatively small distance to
Earth ($d \approx 1.7$~kpc), this is the largest known stellar
association, with a diameter of $\approx 60$~pc and a core radius of
$\sim$10 pc. The cluster age has been
estimated~\cite{Knodlseder:2002eu} from isochrone fitting to be 3--4
Myr, where the age range may reflect a non-coeval star forming event.
The HEGRA experiment has detected an extended TeV $\gamma$-ray source,
J2032+4130, on the outer edge of Cygnus OB2 with no clear counterpart
and a spectrum which can be modelled in terms of either hadronic or
leptonic processes~\cite{Aharonian:2002ij,Aharonian:2005ex}. However
the failure of Chandra and VLA to detect significant levels of X-rays
or radiowaves which would signal the presence of high energy
electrons~\cite{Butt:2003xc} argues for a hadronic mechanism. Above 1
TeV, the HEGRA data can be fitted by a simple power
law~\cite{Aharonian:2005ex}
\begin{widetext}
\begin{equation}
\frac{{\rm d}\Fpho}{{\rm d}\Epho} = 
6.2 \,\,(\pm 1.5_{\rm stat} \pm 1.3_{\rm sys})
\times 10^{-13} \,\, \left(\frac{E_\gamma}{{\rm TeV}}\right)^{-1.9\,
(\pm 0.1_{\rm stat} \pm 0.3_{\rm sys})} \,\,{\rm cm}^{-2} \,\,{\rm
s}^{-1}\,\, {\rm TeV}^{-1} \,\, .
\label{hegra}
\end{equation}
\end{widetext}
The model proposed~\cite{Torres:2003ur} is that protons are
accelerated by the collective effects of stellar winds from massive O
and B stars and only the high energy particles penetrate and interact
in the innermost dense parts of the winds. The colliding protons
generate $\pi^0$'s which produce the observed
$\gamma$-rays. Convection prevents low energy protons from entering
the dense wind region thus explaining the absence of MeV-GeV photons.

At very high energies ($E \gtrsim 10^{8.7}$~GeV) evidence has also
been presented for neutral particles from the Cygnus spiral arm. AGASA
has found a $4\sigma$ correlation of the arrival direction of cosmic
rays at these energies with the Galactic Plane
(GP)~\cite{Hayashida:1998qb}. The GP excess, which is roughly 4\% of
the diffuse flux, is mostly concentrated in the direction of the
Cygnus region~\cite{Teshimaicrc}. Evidence at the 3.2$\sigma$ level
for a GP enhancement at similar energies has also been reported by the
HiRes Collaboration~\cite{Bird:1998nu}. The primary particles must be
neutral (and stable) in order to preserve direction while propagating
through the galactic magnetic field. In principle they can be photons
but this is hard to reconcile with the complete isotropy observed up
to $\sim10^{7.7}$~GeV by KASCADE~\cite{Antoni:2003jm}. Intriguingly,
time-dilated neutrons can reach the Earth from typical Galactic
distances when their energy exceeds $\sim$$10^{9}$~GeV so it is
reasonable to ask whether these might in fact be the primaries.

The GP anisotropy is observed over the energy range $10^{8.9}$ to
$10^{9.5}$~GeV. The lower threshold requires that only neutrons with
energy $\agt 10^{9}$~GeV have a boosted $c\tau_n$ sufficiently large
to serve as Galactic messengers. The decay mean free path of a neutron
is $c\,\Gamma_n\,\overline\tau_n=9.15\,(E_n/10^9~{\rm GeV})$~kpc, the
lifetime being boosted from its rest-frame value,
$\overline\tau_n=886$~seconds, to its lab value by
$\Gamma_n=E_n/m_n$. Actually, the broad scale anisotropy from the
direction of the GP reported by Fly's Eye~\cite{Bird:1998nu} peaks in
the energy bin $10^{8.6} - 10^{9}$~GeV, but persists with less
significance to energies as low as $10^{8.5}$~GeV. This implies that
if neutrons are the carriers of the anisotropy, there {\em needs to
be} some contribution from at least one source closer than $\sim$2
kpc. Interestingly, the full Fly's Eye data includes a directional
signal from the Cygnus region which was somewhat lost in unsuccessful
attempts~\cite{Cassiday:kw,Teshima:1989fc} to relate it to
$\gamma$-ray emission from Cygnus X-3. As shown in Fig.~\ref{cygOB2},
Cygnus OB2 is very close to the line of sight to Cygnus X-3, which is
in fact $\sim8$ kpc farther away than the stellar association.

The upper cutoff reflects an important feature of photodisintegration
at the source: heavy nuclei with energies in the vicinity of the ankle
will fragment to neutrons with energies about an order of magnitude
smaller. To account for the largest neutron energies, it is necessary
to continue the heavy nucleus spectrum to energies above the
ankle~\cite{note1}. Note that the emerging harder extragalactic
spectrum will overwhelm the steeply falling galactic population at
these energies. In order to fit the spectrum in the anisotropy region
and maintain continuity to the ankle region without introducing a
cutoff, the AGASA Collaboration required a spectrum $\propto E_n^{-3}$
or steeper~\cite{Hayashida:1998qb}.

\begin{figure}
\begin{center}
\includegraphics[height=7.8cm]{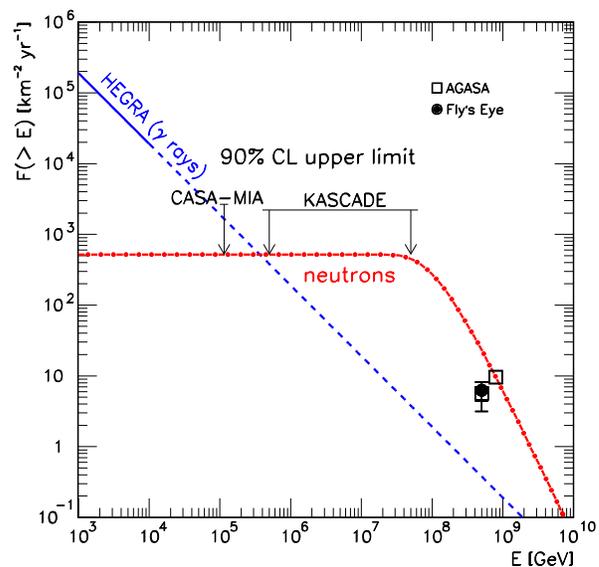}
\caption{The integrated neutron flux expected from Cygnus OB2
(dashed-dotted line) is superimposed over the integrated fluxes
observed from the Cygnus region by the Fly's Eye~\cite{Cassiday:kw}
and AGASA~\cite{Teshima:1989fc,Teshimaicrc} experiments. Also shown is
the $\gamma$-ray flux reported by the HEGRA
experiment~\cite{Aharonian:2005ex} and the upper limits on neutral
particles derived from the CASA-MIA~\cite{Borione:1996jw} and
KASCADE~\cite{Maier:2003av} experiments. The solid line is a fit to
the HEGRA data and the dashed line is the extrapolation to unobserved
energies.}
\label{cygOB2_n}
\end{center}
\end{figure}

In what follows, we model the neutron spectrum with a single power law
reflecting the average shape of the diffuse cosmic ray spectrum
between $10^{6}$ and $10^{8.5}$ GeV, specifically:
\begin{eqnarray}
\frac{{\rm d}F_n}{{\rm d}E_n} & = 
& \frac{{\rm d}F_n}{{\rm d}E_n}\Big|_{\rm source}\,
{\rm e}^{-d/(c\,\Gamma_n\,\overline \tau_n)} \nonumber \\
 & = &
C\, E_n^{-3.1}\,{\rm e}^{- d/(c\,\Gamma_n\,\overline \tau_n)}\,.
\end{eqnarray}
By integrating the spectrum between $E_1 = 10^{8.9}~{\rm GeV}$ and
$E_2 = 10^{9.5}~{\rm GeV}$~\cite{Teshimaicrc}, we can normalize to the
observed integrated flux~\cite{Teshimaicrc}:
\begin{equation}
\int_{E_1}^{E_2} C\, E_n^{-3.1} \,\,\,e^{- d/(c\,\Gamma_n\,\overline \tau_n)}
\,dE_n\, \approx 9~{\rm km}^{-2} {\rm yr}^{-1}\,\,,
\label{integ_n}
\end{equation}
which yields $C=1.15\times 10^{20}~{\rm km}^{-2}~{\rm yr}^{-1}$. We
emphasize again that the neutron primaries hypothesis predicts a
significant signal above the diffuse cosmic ray flux only at energies
$\agt 10^{8.9}$~GeV.  Figure~\ref{cygOB2_n} shows the damping due to
neutron decay which attenuates the directional signal at low energies.
The predicted damped signal for a source at 1.7 kpc is well below
direct limits from the CASA-MIA~\cite{Borione:1996jw} and
KASCADE~\cite{Maier:2003av} experiments.

For every surviving neutron at $\sim 10^9$~GeV, there are many
neutrons at lower energies that do decay via $n\rightarrow p+e^- +
\overline \nu_e.$ The proton is bent by the Galactic magnetic field
and the electron quickly loses energy via synchrotron radiation, but
the $\overline\nu_e$ travels along the initial neutron direction,
producing a directed TeV energy beam which is potentially observable.

The basic formula that relates the neutron flux at the source to the
antineutrino flux observed at Earth is~\cite{Anchordoqui:2003vc}:
\begin{widetext}
\begin{equation}
\frac{{\rm d}\Fnu}{{\rm d}\Enu}(\Enu)  =
\int {\rm d}\En\,\frac{{\rm d}\Fn}{d\En}(\En)\Big|_{\rm source}
\left(1-{\rm e}^{-\frac{D\,m_n}{\En\,\tbar}}\right)\,
\int_0^Q {\rm d}\Enubar\,\frac{{\rm d}P}{{\rm d}\Enubar}(\Enubar)  
\int_{-1}^1 \frac{{\rm d}\cth}{2}
\;\delta\left[\Enu-\En\,\Enubar\,(1+\cth)/m_n\right]
\,.
\label{nuflux}
\end{equation}
\end{widetext}
The variables appearing in Eq.~(\ref{nuflux}) are the antineutrino and
neutron energies in the lab ($\Enu$, $\En$), the antineutrino angle
with respect to the direction of the neutron momentum in the neutron
rest-frame ($\thnu$), and the antineutrino energy in the neutron
rest-frame ($\Enubar$).  The last three variables are not observed by
a laboratory neutrino-detector, and so are integrated over.  The
observable $\Enu$ is held fixed.  The delta-function relates the
neutrino energy in the lab to the three integration variables. The
parameters appearing in Eq.~(\ref{nuflux}) are the neutron mass and
rest-frame lifetime ($m_n$ and $\tbar$). Finally, ${\rm d}P/{\rm
d}\Enubar$ is the normalized probability that the decaying neutron
produces an antineutrino with energy $\Enubar$ in the neutron
rest-frame. Note that the maximum antineutrino energy in the neutron
rest frame is very nearly $Q \equiv m_n - m_p - m_e = 0.71$~MeV.

\begin{figure}
\begin{center}
\includegraphics[height=7.8cm]{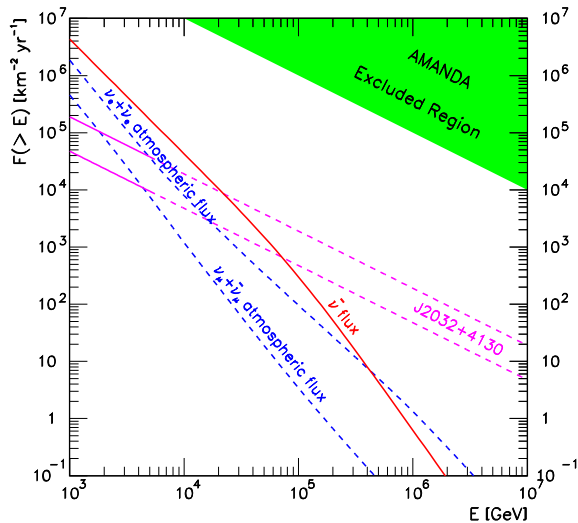}
\caption{Integrated flux of $\overline \nu_\mu + \overline \nu_e +
\overline \nu_\tau$ (solid line) predicted to arrive at Earth from the
direction of the Cygnus region.  Also shown are the integrated
$\nu_\mu + \overline\nu_\mu$ and $\nu_e + \overline \nu_e$ atmospheric
fluxes for an angular bins of $1^\circ \times 1^\circ$ and $10^\circ
\times 10^\circ$, respectively.  The shaded band indicates the region
excluded by the AMANDA experiment~\cite{Ahrens:2003pv}.The fluxes of
neutrinos inferred from HEGRA measurements of the $\gamma$-ray flux
are also shown: the lower line is based on the assumption of $p\gamma$
interactions, whereas the upper line is based on $pp$ interactions
(the charged/neutral pion-production ratio depends on the
interaction). In each case the solid portion of the line indicates the
region where HEGRA data is available and the dashed part is an
extrapolation to unobserved energies.}
\label{cygOB2_nu}
\end{center}
\end{figure}

The integral neutrino flux, $\Fnu(>\Enu)\equiv\int_{\Enu} {\rm
d}\Enu\,\frac{{\rm d}\Fnu}{{\rm d}\Enu}$, is particularly useful for
experiments having a neutrino detection-efficiency that is independent
of neutrino energy, or nearly so.  Our calculated integral flux,
normalized to the integrated neutron flux in Eq.~(\ref{integ_n}), is
shown in Fig.~\ref{cygOB2_nu}.  Note that the nuclear
photodisintegration threshold implies an infrared cutoff on the
primary neutron energy at the source, which in turn leads to a low
energy cutoff of ${\cal O}$(TeV) on the integral flux.

\section{Decoherence Effects in High Energy Neutrino Propagation}
\label{decoherence}

Even though the flux of antineutrinos produced by Cygnus OB2 is pure
$\overline \nu_e,$ the antineutrinos observed at Earth will be
distributed over all flavors. This is because of neutrino
oscillations, as well as possible decoherence effects induced over
long distances (more on this below). In the standard treatment of
neutrino oscillations, neutrino flavor eigenstates,
$|\,\nu_{\alpha}\rangle$, $\alpha=e,\mu,\tau$, are expanded in terms
of mass eigenstates, $|\,\nu_i\rangle$, $i=1,2,3$, through a (unitary)
matrix, $U$, defined by $|\,\nu_{\alpha}\rangle = {\displaystyle
\sum_{i=1}^{3}}U^*_{\alpha i}|\,\nu_i\rangle$. This implies that the
density matrix of a flavor state, $\rho^\alpha $, can be expressed in
terms of mass eigenstates by
$\rho^\alpha=|\nu_\alpha\rangle\langle\nu_\alpha|=\sum_{i,j}U^*_{\alpha
i}U_{\alpha j}|\nu_i\rangle\langle\nu_j| $. This is a pure quantum
system, therefore the density matrix satisfies Tr~$\rho^2$ = Tr~$\rho$
= 1.  To get the transition amplitude, we evolve the system quantum
mechanically with the Liouville equation
\begin{equation}
\frac{\partial \rho}{\partial t} = - i\, [H,\, \rho] \,\,,
\label{liouville}
\end{equation}
where $H$ is the Hamiltonian of the system.  For $\delta m_{ij}^2 d/2E
\gg 1$, the phases induced by the mass splitting, $\delta m_{ij} =
m_i^2-m_j^2$, will be erased by uncertainties in $d$ and $E$, yielding
for the transition probability between flavor states $\alpha$ and
$\beta$~\cite{Learned:1994wg}:
\begin{eqnarray}
P_{\nu_\alpha \to \nu_\beta} & = & {\rm Tr}\, [\rho_\alpha(t) \,
 \rho_\beta] \nonumber \\ & = & \delta_{\alpha \beta} - 2 \sum_{j>i}
 {\rm Re}\, (U_{\beta j}^*\, U_{\beta i}\,U_{\alpha j} \, U_{\alpha i}^*)
 \,\, .
\end{eqnarray}
The prediction for the flavor population at Earth due to standard
flavor-mixing (i.e. with no spacetime dynamics) of the pure
$\overline\nu_e$ source is $\sum_j |U_{ej}|^2 |U_{\alpha j}|^2 \sim
\frac{1}{3} |U_{\alpha 2}|^2 + \frac{2}{3} |U_{\alpha 1}|^2$ for
flavor $\alpha,$ which leads to the flavor ratios $\sim 5:2:2$. This
is very different from the democratic $1:1:1$.

The Hamiltonian evolution in Eq.~(\ref{liouville}) is a characteristic
of physical systems isolated from their surroundings. The time
evolution of such a quantum system is given by the continuous group of
unitarity transformations, $U_t = {\rm e}^{-i H\, t},$ where $t$ is
the time. The existence of the inverse of the infinitesimal generator,
$H,$ which is a consequence of the algebraic structure of the group,
guarantees {\em reversibility} of the processes. For open
quantum-mechanical systems, the introduction of dissipative effects
lead to modifications of Eq.~(\ref{liouville}) that account for the
{\em irreversible} nature of the evolution. The transformations
responsible for the time evolution of these systems are defined by the
operators of the Lindblad quantum dynamical
semi-groups~\cite{Lindblad:1975ef}. Since this does not admit an
inverse, such a family of transformations has the property of being
only forward in time.

The Lindblad approach to decoherence does not require any detailed
knowledge of the environment.  The corresponding time evolution
equation for the density matrix takes the form:
\begin{equation}
\frac{\partial \rho}{\partial t} = - i [H_{\rm eff},\, \rho] + {\cal D}
[\rho] \,\,,
\end{equation}
where the decoherence term is given by
\begin{equation}
{\cal D} [\rho] = - \frac{1}{2} \sum_j \left([b_j,\, \rho\, b_j^\dagger] +
[b_j\, \rho,\, b_j^\dagger]\right)\,\,.
\end{equation}
Here, $H_{\rm eff} = H + H_{\rm d}$ is the effective Hamiltonian of
the system, $H$ is its free Hamiltonian, $H_{\rm d}$ accounts for
possible additional dissipative contributions that can be put in the
Hamiltonian form, and $\{b_j\}$ is a sequence of bounded operators
acting on the
Hilbert space of the open quantum system, ${\cal H}$, and satisfying $\sum_j
b^\dagger_j b_j \in {\cal B} ({\cal H}),$ where ${\cal B} ({\cal H})$
indicates the space of bounded operators acting on ${\cal H}.$

The intrinsic coupling of a microscopic system to the space-time foam
can then be interpreted as the existence of an arrow of time which in
turn makes possible the connection with thermodynamics via an
entropy. The monotonic increase of the von Neumann entropy, $S(\rho) =
- {\rm Tr}\,\, (\rho\, \ln \rho)$, implies the hermiticity of the
Lindblad operators, $b_j = b_j^\dagger$~\cite{Benatti:1987dz}. In
addition, the conservation of the average value of the energy can be
enforced by taking $[H, b_j] = 0$~\cite{Banks:1983by}. The Lindblad
operators of an $N$-level quantum mechanical system can be expanded in
a basis of matrices satisfying standard commutation relations of Lie
groups. For a 3-level system, the basis comprises the eight Gell-Mann
SU(3) matrices plus the $3 \times 3$ identity
matrix~\cite{Adler:2000vf}.

As mentioned above, the theoretical approach provided by Lindblad
quantum dynamical semi-groups is a very general in the sense that no
explicit hypothesis has been made about the actual interactions
causing the loss of coherence. Following Ref.~\cite{Gago:2002na}, we
adopt an expansion in a 3 flavor basis with a diagonal form for the $9
\times 9$ decoherence matrix, ${\cal D}$. Note that neutrinos
oscillate among flavors separately between particle and antiparticle
sectors and so the respective decoherence parameters for antineutrinos
can be different from the corresponding ones in the neutrino
sector. Upon averaging over the rapid oscillation for propagation
between Cygnus OB2 and the Earth, only the diagonal Gell-Mann matrices 
survive, and so  the transition probability for
antineutrinos takes the form~\cite{Gago:2002na}:
\begin{widetext}
\begin{equation}
P_{\overline\nu_\alpha \to \overline\nu_\beta} = \frac{1}{3} +
\left[\frac{1}{2}\,\, {\rm e}^{-\overline\gamma_3 \,d}\,\, (U_{\alpha 1}^2 -
U_{\alpha2}^2) ( U_{\beta 1}^2 - U_{\beta2}^2) + \frac{1}{6}
\,\,{\rm e}^{-\overline \gamma_8 \,d}\,\, (U_{\alpha1}^2 + U_{\alpha 2}^2 -
2 U_{\alpha 3}^2) (U_{\beta 1} + U_{\beta 2}^2 - 2 U_{\beta 3}^2)
\right] \, \, ,
\label{P}
\end{equation}
\end{widetext}
where $\overline \gamma_3$ and $\overline \gamma_8$ are eigenvalues of
the decoherence matrix for antineutrinos. Note that in Eq.~(\ref{P})
we set the $CP$ violating phase to zero, so that all mixing matrix
elements become real. Furthermore, under the assumptions that $CPT$ is
conserved and that decoherence effects are negligible at present
experiments, the values of the mixing angle combinations appearing in
Eq.~(\ref{P}) can be well determined by the usual oscillation analysis
of solar, atmospheric, LBL and reactor
data~\cite{Gonzalez-Garcia:2004jd}. In what follows, we will assume
that $CPT$ is conserved both by neutrino masses and mixing as well as
in decoherence effects. Note however that since the decoherence
effects in the present study primarily affect antineutrinos, our
results will hold for the antineutrino decoherence parameters
exclusively if $CPT$ is violated only through quantum-gravity effects.

Now, we require further $\overline \gamma_3 = \overline \gamma_8\equiv
\overline \gamma$ ( $=\gamma_3 = \gamma_8$ under $CPT$ conservation)
so that Eq.~(\ref{P}) can be rewritten for the case of interest as:
\begin{widetext}
\begin{eqnarray}
&& P_{\overline \nu_e \to \overline \nu_\mu}=
P_{\overline \nu_\mu \to \overline \nu_e}=
P_{\nu_e \to \nu_\mu}=
P_{\nu_\mu \to \nu_e}
= \frac{1}{3} + f_{\nu_e \to \nu_\mu}
{\rm e}^{-\overline\gamma\,d} \,\,,
\nonumber\\
&& P_{\overline \nu_e \to \overline \nu_\tau} =
P_{\overline \nu_\tau \to \overline \nu_e} =
P_{\nu_e \to \nu_\tau} =
P_{\nu_\tau \to  \nu_e}
= \frac{1}{3} + f_{\nu_e \to \nu_\tau} 
{\rm e}^{-\overline\gamma\,d} \,\,,
\nonumber
\\
&& P_{\overline \nu_\mu \to \overline \nu_\tau}=
P_{\overline \nu_\tau \to \overline \nu_\mu}=
P_{\nu_\mu \to \nu_\tau}=
P_{\nu_\tau \to  \nu_\mu}
= \frac{1}{3} + f_{\nu_\mu \to \nu_\tau}
{\rm e}^{-\overline\gamma\,d} \,\,, \label{eq:probdeco}\\
&&
P_{\overline \nu_e \to \overline \nu_e} =
P_{\nu_e \to \nu_e} =
\frac{1}{3}-(f_{\nu_e \to \nu_\mu}+f_{\nu_e \to \nu_\tau})\,
{\rm e}^{-\overline\gamma\,d} \,\,,\nonumber\\
&&
P_{\overline \nu_\mu \to \overline \nu_\mu} =
P_{\nu_\mu \to \nu_\mu} =
\frac{1}{3}-(f_{\nu_e \to \nu_\mu}+f_{\nu_\mu \to \nu_\tau})\,
{\rm e}^{-\overline\gamma\,d} \,\,,\nonumber\\
&&
P_{\overline \nu_\tau \to \overline \nu_\tau} =
P_{\nu_\tau \to \nu_\tau} =
\frac{1}{3}-(f_{\nu_e \to \nu_\tau}+f_{\nu_\mu \to \nu_\tau})\,
{\rm e}^{-\overline\gamma\,d}\,\,. \nonumber
\end{eqnarray}
\end{widetext}

We make this simplification only to emphasize the primary signature of
quantum decoherence, namely that after travelling a sufficiently long
distance the flavor mixture is altered to the ratio $1:1:1,$
regardless of the initial flavor content. Consequently, if a flux of
antineutrinos were to be observed from the Cygnus spiral arm with a
flavor ratio $\neq 1:1:1,$ strong constraints can be placed on the
decoherence parameter $\overline \gamma$.

Using the results of the up-to-date 3-$\nu$ oscillation analysis of
solar, atmospheric, LBL and reactor data~\cite{Gonzalez-Garcia:2004jd}
we obtain the following values and 95\% confidence ranges
\begin{eqnarray}
f_{\nu_e \to  \nu_\mu}&=&-0.106^{+0.060}_{-0.082} \,\,,\nonumber\\
f_{\nu_e \to  \nu_\tau}&=&-0.128^{+0.089}_{-0.055} \,\,, \label{fit} \\
f_{\nu_\mu \to  \nu_\tau}&=&\phantom{-}0.057^{+0.011}_{-0.035} \,\, . \nonumber
\end{eqnarray}
The numbers given in Eq.~(\ref{fit}) are obtained using the same
techniques as described in Ref.~\cite{Gonzalez-Garcia:2004jd} but
including the final SNO salt phase data~\cite{Miknaitis:2005rw}.

We assume a phenomenological parametrization for the eigenvalues of
the decoherence matrix~\cite{Lisi:2000zt},
\begin{equation}
\overline\gamma =
\kappa_n \,\, (E_\nu/{\rm GeV})^n,
\label{kk}
\end{equation}
with the integer $n \in [-1,\, 3]$. This allows a straightforward
comparison with existing limits.  Equipped with
Eqs.~(\ref{eq:probdeco}), (\ref{fit}) and~(\ref{kk}), we now proceed
to determine the sensitivity of IceCube to decoherence effects.

\section{Sensitivity Reach at ICECUBE}
\label{sec:reach}

In deep ice, neutrinos are detected through the observation of \v
{C}erenkov light emitted by charged particles produced in neutrino
interactions. In the case of an incident high-energy muon neutrino,
for instance, the neutrino interacts with a hydrogen or oxygen nucleus
in the deep ice and produces a muon traveling in nearly the same
direction as the neutrino. The blue \v {C}erenkov light emitted along
the muon's kilometer-long trajectory is detected by strings of
PhotoMultiplier Tubes (PMTs) deployed at depth shielded from
radiation. The orientation of the \v {C}erenkov cone reveals the
neutrino direction. There may also be a visible hadronic shower if the
neutrino is of sufficient energy.

The Antarctic Muon And Neutrino Detector Array
(AMANDA)~\cite{Andres:1999hm}, using natural 1 mile deep Antarctic ice
as a \v {C}erenkov detector, has operated for more than three years in
its final configuration: 19 strings instrumented with 680 PMTs.
IceCube~\cite{Ahrens:2003ix}, the successor experiment to AMANDA, is
now under construction. It will consist of 80 kilometer-length
strings, each instrumented with 60 PMTs spaced by 17~m.  The deepest
module is 2.4~km below the surface. The strings are arranged at the
apexes of equilateral triangles 125\,m on a side. The instrumented
(not effective!)  detector volume is a full cubic kilometer. A surface
air shower detector, IceTop, consisting of 160
Auger-style~\cite{Abraham:2004dt} \v {C}erenkov detectors deployed
over 1\,km$^{2}$ above IceCube, augments the deep-ice component by
providing a tool for calibration, background rejection and air-shower
physics.  The expected energy resolution is $\pm 0.1$ on a log$_{10}$
scale.  Construction of the detector started in the Austral summer of
2004/2005 and will continue for 6 years, possibly less. At the time of
writing, data collection by the first string has begun.

At IceCube, the events are grouped as either muon tracks or
showers. Tracks include muons resulting from both cosmic muons and
from Charged Current (CC) interaction of muon neutrinos. The angular
resolution for muon tracks $\approx 0.7^\circ$~\cite{Ahrens:2002dv}
allows a search window of $1^\circ \times 1^\circ$.  This corresponds
to a search bin solid angle of $\Delta \Omega_{1^\circ \times 1^\circ}
\approx 3 \times 10^{-4}$~sr. In order to reduce the background from
cosmic muons, we adopt here the quality cuts referred to as ``level 2"
cuts in Ref.~\cite{Ahrens:2003ix}.

In our semianalytical calculation~\cite{Gonzalez-Garcia:2005xw}, we
estimate the expected number of $\nu_\mu$ induced tracks from the
Cygnus OB2 source antineutrino flux as
\begin{widetext}
\begin{equation}
{\cal N}_{\rm S}^{\rm tr}
 = T \, n_{\rm T}\,
\int^\infty_{l'_{\rm min}} {\rm d}l\,
\int_{m_\mu}^\infty {\rm d}E_\mu^{\rm fin}\,
\int_{E_\mu^{\rm fin}}^\infty {\rm d}E_\mu^0\,
\int_{E_\mu^0}^\infty {\rm d}E_{\overline \nu}
\frac{{\rm d} F_{\overline \nu_\mu}}{{\rm d}E_{\overline \nu}}(E_{\overline \nu})
\frac{{\rm d}\sigma_{\rm CC}}{{\rm d}E_\mu^0}(E_{\overline \nu},E_\mu^0)\,
F(E^0_\mu, E_\mu^{\rm fin}, l)\, A^0_{\rm eff}\, \,,
\label{eq:trsource}
\end{equation}
\end{widetext}
where
\begin{equation}
\frac{{\rm d} F_{\overline\nu_\alpha}}{{\rm d}E_{\overline \nu}}
= P_{\overline\nu_e\to\overline\nu_\alpha}(E_{\overline \nu})
\frac{{\rm d} F_{\overline \nu}}{{\rm d}E_{\overline \nu}}
\end{equation}
is the differential antineutrino flux which arrives at the Earth with
flavour $\alpha$ after oscillation of the $\overline \nu_e$ in
Eq.(\ref{nuflux}). Here $\frac{{\rm d}\sigma_{\rm CC}}{{\rm
d}E_\mu^0}(E_{\overline\nu},E_\mu^0)$ is the differential CC
interaction cross-section producing a muon of energy $E_\mu^0$,
$n_{\rm T}$ is the number density of nucleons in the matter
surrounding the detector, and $T$ is the exposure time of the
detector. After being produced, the muon propagates through the rock
and ice surrounding the detector and loses energy. We denote by
$F(E^0_\mu,E_\mu^{\rm fin},l)$ the function that represents the
probability of a muon produced with energy $E_\mu^0$, arriving at the
detector with energy $E_\mu^{\rm fin}$, after traveling a distance,
$l$. The details of the detector are encoded in the effective area,
$A^0_{\rm eff}$. We use the parametrization of the $A^0_{\rm eff}$
described in Ref.~\cite{Gonzalez-Garcia:2005xw} to simulate the
response of the IceCube detector after events that are not neutrinos
have been rejected (this is achieved by quality cuts referred to as
``level 2'' cuts in Ref.~\cite{Ahrens:2003ix}). The minimum track
length cut is $l_{\rm min}=300$~m and we account for events with
$E_\mu^{\rm fin}>500$ GeV. With this we find that, assuming standard
neutrino oscillations, one expects a total of $212\times
P_{\overline\nu_e\to\overline\nu_\mu}=48$ $\overline\nu_\mu$-induced
tracks from the Cygnus OB2 source in 15 years of observation.

Showers are generated by neutrino collisions --- $\nu_e\ \mbox{or}\,\,
\nu_\tau$ CC interactions, and all Neutral Current (NC) interactions
--- inside of or nearby the detector, and by muon bremsstrahlung
radiation near the detector. For showers, the angular resolution is
significantly worse than for muon tracks. In our analysis, we
consider a shower search bin solid angle, $\Delta \Omega_{10^\circ
\times 10^\circ}.$ Normally, a reduction of the muon bremsstrahlung
background is effected by placing a cut of 40 TeV on the minimum
reconstructed energy~\cite{Ackermann:2004zw}. For Cygnus OB2, this
strong energy cut is not needed since this background is filtered by
the Earth. Thus we account for all events with shower energy $E_{\rm
sh}\geq E_{\rm sh}^{\rm min}=1$~TeV. The directionality requirement,
however, implies that the effective volume for detection of showers is
reduced to the instrumented volume of the detector, 1 ${\rm km}^3$,
because of the small size of the showers (less than 200 m in radius)
in this energy range.

We can now calculate the expected number of showers from the Cygnus
OB2 source to be:
\begin{equation}
{\cal N}_{\rm S}^{\rm sh}  = 
{\cal N}_{\rm S}^{\rm sh,CC}+{\cal N}_{\rm S}^{\rm sh,NC} \,\,,
\end{equation}
where
\begin{widetext}
\begin{equation}
{\cal N}_{\rm S}^{\rm sh,CC} =  T\, n_{\rm T}\, {\cal V}_{\rm eff}\,
\int_{E^{\rm min}_{\rm sh}}^\infty {\rm d}E_{\overline \nu} \, \sum_{\alpha=e,\tau}\frac
{{\rm d} F_{\overline \nu_\alpha}}{{\rm d}E_{\overline \nu}}(E_{\overline \nu})
\sigma_{\rm CC}(E_{\overline \nu}) \,\,,
\label{eq:shsourcecc}
\end{equation}
and
\begin{equation}
{\cal N}_{\rm S}^{\rm sh,NC} =  T\, n_{\rm T}\, {\cal V}_{\rm eff}\,
\int_{E_{\overline\nu}-E^{\rm min}_{\rm sh}}^\infty {\rm d}E'_{\overline \nu}
\int_{E^{\rm min}_{\rm sh}}^\infty {\rm d}E_{\overline \nu} \, \sum_{\alpha=e,\mu,\tau}
\frac{{\rm d} F_{\overline \nu_\alpha}}{{\rm d}E_{\overline \nu}}(E_{\overline \nu})
\frac{{\rm d}\sigma_{\rm NC}}{{\rm d}E'_{\overline\nu}}
(E_{\overline \nu}, E'_{\overline \nu}) \,\,.
\label{eq:shsourcenc}
\end{equation}
\end{widetext}
Here, $\frac{{\rm d}\sigma_{\rm NC}}{{\rm d}E'_{\overline\nu}}
(E_{\overline \nu},E'_{\overline \nu})$ is the differential NC
interaction cross section producing a secondary antineutrino of
energy, $E'_{\overline\nu}$. In writing Eqs.~({\ref{eq:shsourcecc}})
and (\ref{eq:shsourcenc}) we are assuming that for contained events
the shower energy corresponds with the interacting $\overline\nu_e$ or
$\overline\nu_\tau$ antineutrino energy ($E_{\rm
sh}=E_{\overline\nu}$) in a CC interaction, while for NC the shower
energy corresponds to the energy in the hadronic shower $E_{\rm
sh}=E_{\overline\nu}-E'_{\overline\nu}\equiv E_{\overline\nu}\, y,$
where $y$ is the usual inelasticity parameter in DIS. In total, within
the framework of standard oscillations, we expect 25 showers from the
Cygnus OB2 source in 15 years of operation.

We now turn to the estimate of the background. There are two different
contributions --- atmospheric neutrinos and additional fluxes of
extraterrestrial neutrinos. For the ``conventional'' atmospheric
neutrino fluxes arising from pion and kaon decays, we adopt the
3-dimensional scheme estimates of Ref.~\cite{Honda:2004yz}, which we
extrapolate to match at higher energies the 1-dimensional calculations
by Volkova~\cite{Volkova:1980sw}. We also incorporate ``prompt''
neutrinos from charm decay as calculated in
Ref.~\cite{Gondolo:1995fq}.  We obtain the number of expected track
and shower events from atmospheric neutrinos as in
Eqs~(\ref{eq:trsource}), (\ref{eq:shsourcecc}), and
(\ref{eq:shsourcenc}) with $\frac{{\rm d} F^{\rm
ATM}_{\nu_\alpha}}{{\rm d}E_{\nu}}(E_{\nu})$ being the $\nu_e$ and
$\nu_\mu$ atmospheric neutrino fluxes integrated over a solid angle of
of $1^\circ\times 1^\circ$ (for tracks) and $10^\circ\times 10^\circ$
(for showers) width around the direction of the Cygnus OB2 source
($\theta=131.2^\circ$). We get an expected background of 14
atmospheric tracks and 47 atmospheric showers in 15 years. Of the 47
showers, 16 correspond to $\nu_e$ CC interactions while 31 correspond
to $\nu_\mu$ NC interactions. The large yield of NC events is due to
the fact that at these energies, the atmospheric flux contains a very
unequal mix of neutrino flavors (with ratios $\approx 1:20:0$). We
have also verified that if we increase the minimum shower energy cut
to 5 TeV, $\nu_e$ CC and $\nu_\mu$ NC contribute in equal amounts to
the number of atmospheric showers. This is in agreement with
simulations by the AMANDA Collaboration~\cite{Ahrens:2002wz}.

We turn now to the discussion of background events from other
extraterrestrial sources.  As discussed in Sec.~\ref{neutrons}, the
TeV $\gamma$-ray flux reported by the HEGRA
Collaboration~\cite{Aharonian:2005ex} in the vicinity of Cygnus OB2 is
likely to be due to hadronic processes: the $\gamma$-rays can be
directly traced to the decay $\pi^0$'s produced through inelastic $pp$
collisions~\cite{Torres:2003ur}. Since $\pi^0$'s, $\pi^+$'s, and
$\pi^-$'s are produced in equal numbers, we expect two photons, two
$\nue$'s, and four $\numu$'s per $\pi^0$. On average, the photons
carry one-half of the energy of the pion, and the neutrinos carry
one-quarter. During propagation, the $\numu$'s will partition
themselves equally between $\numu$'s and $\nutau$'s on lengths large
compared to the oscillation length and so one finds at Earth a nearly
identical flux for the three neutrino flavors~\cite{Anchordoqui:2004eu}:
\begin{equation}
\frac{{\rm d}F_{\nu_\alpha}}{{\rm d}E_\nu} (E_\nu = \Epho/2)
= 2\,\frac{{\rm d}\Fpho}{{\rm d}\Epho}(\Epho)\,.
\label{oneoneone}
\end{equation}
For $p \gamma \to N \pi$ interactions, it can easily be shown using
the $\Delta$-approximation that the resulting neutrino flux is about a
factor of 4 smaller~\cite{Anchordoqui:2004eb}. For the purposes of
setting an upper bound on the neutrino flux we ignore all other
sources near J2032+4130 because their steady emission in $\gamma$-rays
is estimated to be smaller by more over a factor of 5 than the source
of interest~\cite{cygnusx3}. Substituting Eq.~(\ref{hegra}) into
Eq.~(\ref{oneoneone}) we obtain the corresponding background from
neutrinos with flavor ratios 1:1:1. As can be seen in
Fig.~\ref{cygOB2_nu}, the background is dominated by atmospheric
neutrinos. Thus after 15~years of data collection we expect 18 tracks
and 1 shower from J2032+4130 for standard oscillations. In summary,
the directional beam from the Cygnus region provides a statistically
significant signal to probe anomalous oscillations in the antineutrino
sector.

We will now discuss how to isolate the possible signal due to
decoherence in the antineutrinos from Cygnus OB2 from the atmospheric
background and possible fluctuations in the event rate due to unknown
diffuse fluxes of extraterrestrial neutrinos.  In general, we can
predict that the expected number of track and shower events in the
direction of the Cygnus OB2 source to be
\begin{eqnarray}
{\cal N}^{\rm tr} &=& {\cal N}_{\rm S}^{\rm tr} + {\cal N}_{\rm ATM}^{\rm tr} +
  {\cal N}_{\rm SS}^{\rm tr} \,\,,
\label{pepetr}\\
{\cal N}^{\rm sh} &=& {\cal N}_{\rm S}^{\rm sh} + {\cal N}_{\rm ATM}^{\rm sh} +
  {\cal N}_{\rm SS}^{\rm sh}.
\label{pepesh}
\end{eqnarray}
The first term corresponds to antineutrinos from neutron
$\beta$-decay. In the presence of decoherence effects these event
rates can be computed from
Eqs.~(\ref{eq:trsource}),~(\ref{eq:shsourcecc})~and~(\ref{eq:shsourcenc})
with flavour transition probabilities given in Eq.~(\ref{eq:probdeco})
with $d=1.7$ kpc.  The second term refers to atmospheric
(anti)neutrinos (${\cal N}_{\rm ATM}^{\rm tr}=14$, ${\cal N}_{\rm
ATM}^{\rm sh}=47$ for 15 years of exposure). The third term takes into
account additional contributions from a diffuse flux of
(anti)neutrinos produced via charged pion decay. In principle,
decoherence effects may also affect the expected number of events from
this diffuse flux. However given that the flavour ratios both from
oscillation and decoherence are very close to 1:1:1 for the case of
neutrinos produced via charge pion decay, we find that there is no
difference in the sensitivity region if decoherence effects are
included or not in the evaluation of $ {\cal N}_{\rm SS}^{\rm tr}$ and
$ {\cal N}_{\rm SS}^{\rm sh} $.  They are normalized to the maximum
expected flux from J2032+4130 by a factor $x= {\cal N}_{\rm SS}^{\rm
tr}/18= {\cal N}_{\rm SS}^{\rm sh}/1$.

Altogether, the quantities ${\cal N}^{\rm tr}$ and ${\cal N}^{\rm
sh}$, as defined in Eqs.~(\ref{pepetr}) and (\ref{pepesh}), can be
regarded as the theoretical expectations of these events rates,
corresponding to different points in the $x-\kappa_n$ parameter space.
For a given set of observed rates, ${\cal N}_{\rm obs}^{\rm tr}$ and
${\cal N}_{\rm obs}^{\rm sh}$, two curves are obtained in the
two-dimensional parameter space by setting ${\cal N}_{\rm obs}^{\rm
tr}={\cal N}^{\rm tr}$ and ${\cal N}_{\rm obs}^{\rm sh}={\cal N}^{\rm
sh}.$ These curves intersect at a point, yielding the most probable
values for the flux and decoherence scale for the given observations.
Fluctuations about this point define contours of constant $\chi^2$ in
an approximation to a multi-Poisson likelihood analysis. The contours
are defined by~\cite{Baker:1983tu}
\begin{equation}
\chi^2 = \sum_i^{{\rm sh},\, {\rm tr}} 2 \left[{\cal N}^i - {\cal
N}_{\rm obs}^i + {\cal N}_{\rm obs}^i \, \ln\left(\frac{{\cal N}_{\rm
obs}^i}{{\cal N}^i} \right)\right] \,\,.
\end{equation}

\begin{figure}
\begin{center}
\includegraphics[height=7.8cm]{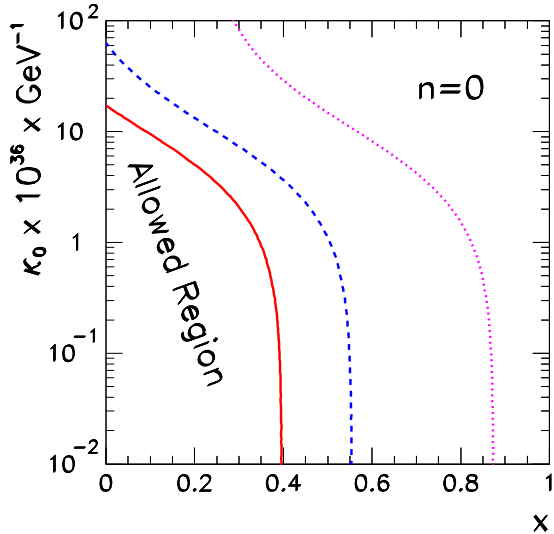}
\caption{IceCube's sensitivity to quantum decoherence assuming ${\cal
N}^{\rm tr}_{\rm obs} = 62$ tracking events and ${\cal N}^{\rm
sh}_{\rm obs} = 72$ showering events from Cygnus. The regions above
and to the right of the solid, dashed and dotted lines can potentially
be excluded at 90\%, 95\%, and 99\% confidence level, respectively.}
\label{iceqg1}
\end{center}
\end{figure}

As illustration, in Fig.~\ref{iceqg1} we show for the case $n=0$, the
expected constraints on $\kappa_0$ at 90, 95 and 99\% CL for 2
d.o.f. if observations turn out to be in agreement with standard
neutrino oscillation expectations, taking ${\cal N}^{\rm tr}_{\rm obs}
=62 $ and ${\cal N}^{\rm sh}_{\rm obs} = 72$ (and no diffuse
flux). Similar regions can be obtained for other choices of $n$.

Marginalizing with respect to $x$, we extract the following 1
degree-of-freedom bounds on the decoherence parameters
\begin{eqnarray}
\kappa_{-1}&\leq& 1.0 \times 10^{-34} \;\; (2.3\times 10^{-31}) \;{\rm GeV}\\
\kappa_{0}&\leq& 3.2\times 10^{-36} \;\; (3.1\times 10^{-34}) \;{\rm GeV}\\
\kappa_{1}&\leq &1.6\times 10^{-40} \;\; (7.2\times 10^{-39}) \;{\rm GeV}\\
\kappa_{2}&\leq& 2.0\times 10^{-44} \;\; (5.5\times 10^{-42}) \;{\rm GeV}\\
\kappa_{3}&\leq& 3.0\times 10^{-47} \;\; (2.9\times 10^{-45}) \;{\rm GeV}
\end{eqnarray}
at 90 (99) \% CL. These should be compared with the current 90\% CL
upper limits on the decoherence parameter from the Super-Kamiokande
and K2K data: $\kappa_{-1}\leq 2.0 \times 10^{-21}$~GeV~,
$\kappa_0\leq 3.5 \times 10^{-23}$~GeV~ and $\kappa_2\leq 9.0 \times
10^{-28}$~GeV~\cite{Lisi:2000zt}. It is clear that IceCube will
provide a major improvement in sensitivity to the possible effects of
quantum gravity.

\section{Discussion}
\label{discussion}

Having demonstrated that IceCube will be able to set bounds on quantum
decoherence effects well beyond the levels currently probed, we now
comment briefly on the theoretical implications.

Any type of high energy/short distance space-time foam interaction
given in Eq.~(\ref{P}) can be understood in analogy with the tracing
out of the degrees of freedom of a thermal bath (with temperature $T$)
with which the open system (in our case a neutrino beam) interacts. A
simple version of the interaction with the bath can be written as
$H_{\rm int}= \sum_j b_j (a+ a^\dagger),$ where $a,\; a^\dagger$ are
raising and lowering operators for space-time foam excitations, with
$\langle a^\dagger a\rangle = (e^{E_{\rm bath}/T} -
1)^{-1}$~\cite{Davydov}.

The energy behavior of $\overline\gamma$ depends on the dimensionality
of the operators $b_j.$ But care must be taken, since ${\cal D}$ is
bilinear in the $b_j,$ and due to the hermiticity requirement, each
$b_j$ is itself at least bilinear in the neutrino fields
$\psi$. Examples are
\begin{equation}
b_j \propto \int d^3x\,\, \psi^\dagger\,\, (i\partial_t)^j \psi\ \ ,
\label{bj}
\end{equation}
where $j=0,\,1,\,2\dots .$ A Fourier expansion of the fields
$\psi,\,\psi^\dagger$, inserted into Eq.~(\ref{bj}), gives the energy
behavior $b_j\propto E_{\overline \nu}^j,$ and hence
$\overline\gamma\propto E_{\overline \nu}^{2j}.$ This restriction of
the energy behavior to non-negative even powers of $E_{\overline \nu}$
may possibly be relaxed when the dissipative term is directly
calculated in the most general space-time foam background.

An interesting example is the case where the dissipative term is
dominated by the dimension-4 operator $b_1,$ $\int d^3x\,
\psi^\dagger\, i\partial_t \psi\ ,$ yielding the energy dependence
$\overline\gamma \propto E_{\overline \nu}^{2}/M_{\rm Pl}.$ This is
characteristic of non-critical string theories where the space-time
defects of the quantum gravitational ``environment'' are taken as
recoiling $D$-branes, which generate a cellular structure in the
space-time manifold~\cite{Ellis:1996bz}.

Although the cubic energy dependence $\overline\gamma\propto
E_{\overline \nu}^3$ is not obtainable from the simple operator
analysis presented above, it may be heuristically supported by a
general argument that each of the $b_j$ must be suppressed by at least
one power of $M_{\rm Pl},$ giving a leading behavior
\begin{equation}
\overline\gamma = \tilde\kappa\ E_{\overline \nu}^{3}/ M_{\rm Pl}^2\,\, .
\end{equation}
Here $\tilde\kappa$ is a dimensionless parameter which by naturalness
is expected to be ${\cal O}(1).$ Decoherence effects with this energy
behavior are undetectable by existing experiments. However, since the
loss of quantum coherence is weighted by the distance travelled by the
antineutrinos, by measuring the $\overline \nu$-Cygnus beam IceCube
will attain a sensitivity down to $\tilde\kappa \lesssim 3.0 \times
10^{-7}$ at 99\% CL, well below the natural expectation.

Finally, we note an interesting aspect of the $\kappa_{-1}$ limit.
For $n=-1$, a non-vanishing $\overline \gamma$ in
Eq.~(\ref{eq:probdeco}) can be related to a finite $\overline \nu_e$
lifetime in the lab system~\cite{Barger:1998xk}:
\begin{equation}
{\rm e}^{-\overline \gamma \,d} \equiv {\rm e}^{-d/\tau_{\rm lab}} =
{\rm e}^{-d\,m_{\overline\nu_e}/E_{\overline \nu}\, \overline
\tau_{\overline\nu_e}}\,\,,
\end{equation}
where $\overline \tau_{\overline\nu_e}$ is the antineutrino rest frame
lifetime and $m_{\overline\nu_e}$ its mass. Therefore the 90\% bound
from IceCube on $\kappa_{-1}$ can be translated into
\begin{equation}
\frac{\overline \tau_{\overline\nu_e}} {m_{\overline\nu_e}}> 10^{34}~{\rm GeV}^{-2} \equiv 6.5\,\,
{\rm s}~{\rm eV}^{-1}\, \,\, .
\label{mtau}
\end{equation}
This corresponds to an improvement of about 4 orders of magnitude over
the best existing bounds from solar neutrino
data~\cite{Beacom:2002cb}, and of course gives results comparable to
the reach derived previously for neutrinos decaying over cosmic
distances~\cite{Beacom:2002vi}.  It should be noted that although the
similar algebraic structure of the decoherence term in
Eq.~(\ref{eq:probdeco}) and a decaying component in the neutrino beam
provide a bound on the neutrino lifetime, these are conceptually two
different processes. The decoherence case can be viewed as a
successive absorption and re-emission of a neutrino from the quantum
foam with change in flavor but no change in the average energy because
of the condition $[H,b_j]=0$. This contrasts with the decay process
which involves the emission of an additional particle.

In conclusion, the IceCube experiment will be sensitive to the effects
of quantum decoherence at a level well beyond current limits. In
particular, antineutrinos produced in the decays of neutrons from
Cygnus OB2 provide an excellent source in which to search for such
effects~\cite{Briceno:2004ik}. Although the precise conclusions depend
on the model considered, we find that in general IceCube can achieve 
a sensitivity of more than 10 orders of magnitude beyond current
bounds on decoherence through observations of Galactic sources of
neutrons.

\acknowledgments{We would like to thank Felix Aharonian, Gavin Rowell,
and the HEGRA Collaboration for permission to reproduce
Fig.~\ref{cygOB2}.  LAA and HG are supported by the US National
Science Foundation (NSF) Grants Nos. PHY-0457004 and PHY-0244507,
respectively.  MCG-G is supported by the US NSF Grant No. PHY0354776
and by Spanish Grants FPA-2004-00996 and GRUPOS03/013.  FH is
supported in part by the US NSF under Grant No. OPP- 0236449, in part
by the US Department of Energy (DoE) Grant No. DE-FG02-95ER40896, and
in part by the University of Wisconsin Research Committee with funds
granted by the Wisconsin Alumni Research Foundation. DH is supported
by the Leverhulme trust. SS is supported by PPARC. TJW is supported 
by the DoE DE-FG05-85ER40226 and NASA-ATP 02-000-0151 in the USA, and 
by the Center for Fundamental Physics (CfFP) at CCLRC Rutherford-Appleton 
Laboratory, UK.}


\begin{thebibliography}{99}
\bibitem{note1} To produce the highest energy neutrons (with energy
$\alt 10^{9.2}$~GeV) via photodisintegration of medium mass (say, $A=
10 - 20$) nuclei, one needs primary particles with energies $\alt
10^{10.2}$~GeV. For a falling spectrum $\propto E_{\rm CR}^{-3},$ the
medium mass nucleus population at $10^{10.2}$~GeV is roughly 3 orders
of magnitude smaller than the diffuse flux at $10^{9.2}$~GeV. This
means that it is about 1.6 orders of magnitude smaller than the
reported cosmic ray excess. However, since each nucleus produces
roughly $A-Z$ neutrons, the average number of liberated neutrons is on
the same order of magnitude, $1.6 - \log_{10} (A-Z)$, than the obseved
neutron population at $\alt 10^{9.2}$~GeV.
\end{thebibliography}
\end{document}